\documentclass[a4paper]{article}

\usepackage{INTERSPEECH2022}
\usepackage{multirow}
\usepackage{enumitem}
\usepackage{cite}

\def\ItemizeZeroPad{\setlength{\itemsep}{0pt}\setlength{\parskip}{0pt}\setlength{\parsep}{0pt}}

\def\VocabSampled{\mathcal{V}^\mathsf{sampled}}
\def\VocabPos{\mathcal{V}^\mathsf{pos}}
\def\VocabNeg{\mathcal{V}^\mathsf{neg}}
\def\TransducerLoss{\mathcal{L}_\mathsf{transducer}}
\def\Blank{\langle\texttt{blank}\rangle}

\title{Memory-Efficient Training of RNN-Transducer with Sampled Softmax}
\name{Jaesong Lee$^{1*}\thanks{* Two authors contributed equally.}$, Lukas Lee$^{1*}$, Shinji Watanabe$^2$}
\address{
  $^1$Naver Corporation\\
  $^2$Carnegie Mellon University}
\email{jaesong.lee@navercorp.com, lukas.lee@navercorp.com, shinjiw@ieee.org}

\begin{document}

\maketitle
\begin{abstract}
RNN-Transducer has been one of promising architectures for end-to-end automatic speech recognition.
Although RNN-Transducer has many advantages including its strong accuracy and streaming-friendly property,
its high memory consumption during training has been a critical problem for development.
In this work, we propose to apply sampled softmax to RNN-Transducer, which requires only a small subset of vocabulary during training thus saves its memory consumption.
We further extend sampled softmax to optimize memory consumption for a minibatch, and employ distributions of auxiliary CTC losses for sampling vocabulary to improve model accuracy.
We present experimental results on LibriSpeech, AISHELL-1, and CSJ-APS,
where sampled softmax greatly reduces memory consumption and still maintains the accuracy of the baseline model.
\end{abstract}

\section{Introduction}
\label{sec:introduction}

End-to-end automatic speech recognition (ASR) has been considered significant over the past few years~\cite{graves2012sequence,graves2013speech,hannun2014deep,chorowski2015attention,miao2015eesen,amodei2016deep,chan2016listen,kim2017joint,chiu2018state,chiu2018monotonic,baevski2020wav2vec}.
There have been large improvements in the performance of end-to-end ASR with advances in deep learning approaches and frameworks. Furthermore, training and inference pipelines of end-to-end ASR are relatively simple compared to those of traditional hidden Markov
model (HMM) based methods. As a result, end-to-end ASR has been a promising direction from research to production~\cite{rao2017exploring,guo2020efficient}.

RNN-Transducer~\cite{graves2012sequence} is one of end-to-end ASR architectures that has been in the spotlight~\cite{rao2017exploring,gulati2020conformer,battenberg2017exploring,li2019improving,he2019streaming,guo2020efficient}.
It consists of an encoder network, a prediction network and a joint network in a single model.
Acoustic features are processed in the encoder network and text history is considered in the prediction network.
RNN-Transducer has several advantages for production.
Its auto-regressive characteristic makes it easier to deal with language model (LM) information of target symbols,
compared to Connectionist Temporal Classification (CTC)~\cite{graves2006connectionist}.
Also, RNN-Transducer is streaming-friendly and often outperforms conventional encoder-decoder or monotonic attention architectures for long-form speech~\cite{chiu2019comparison,kim2021comparison}.

However, the joint network and the loss function of RNN-Transducer consume significantly large memory during training,
thus, the development of RNN-Transducer can be computationally expensive compared to other end-to-end architectures.
This prevents developing RNN-Transducer without large GPU clusters.
Various ways are developed to mitigate the problem~\cite{wang2019exploring,burchi2021efficient,wang2021cascade,saon2021advancing,li2019improving,graves2013speech,rao2017exploring,hu2020exploring,panchapagesan2021efficient},
but most of them require language-specific or network-specific engineering efforts.
% , including modification of network~\cite{wang2019exploring,burchi2021efficient,saon2021advancing},
% language-dependent engineering~\cite{wang2021cascade}, and pre-training of network~\cite{graves2013speech,rao2017exploring,hu2020exploring}.

In this work, we propose a new way to train RNN-Transducer without large memory consumption.
We apply sampled softmax~\cite{jean2015using}, which is originally developed for neural machine translation, to RNN-Transducer.
For computing the RNN-Transducer loss, only a small subset of vocabulary is sampled and used for the softmax function, so the memory bottleneck can be reduced by adjusting the size of the subset during training.

To efficiently train RNN-Transducer models, we extend sampled softmax in two ways.
First, we break the common assumption of sampled softmax that a sampled subset is shared within minibatch.
Instead, we propose to sample the subset independently per each training example,
which gives a significant reduction of the subset size and thus saves more memory.

Second, we propose to use a token posterior distribution of joint CTC loss~\cite{jeon2021multitask,boyer2021study} for sampling the subset.
We show that a choice of a sampling distribution affects the model accuracy, and that a joint CTC distribution provides good accuracy while requiring only marginal computational costs.
We also apply Intermediate CTC~\cite{lee2021intermediate} and Self-conditioned CTC~\cite{jumon2021relaxing} to RNN-Transducer, which are both used for regularization of a model and for the sampling distribution of sampled softmax.
With the two extensions of sampled softmax, the RNN-Transducer model can be trained with much less memory consumption and reaches the accuracy of the original model.

In summary, we show the following contributions:
\begin{itemize}[leftmargin=*]\ItemizeZeroPad
\item We propose to apply sampled softmax to RNN-Transducer for memory-efficient training.
\item We extend sampled softmax with the example-wise sampling strategy and with use of a token distribution of joint CTC branch for a sampling distribution.
\item We extend Intermediate CTC and Self-conditioned CTC to RNN-Transducer, and also employ them for sampling distributions of sampled softmax.
\item We experimentally show that sampled softmax greatly reduces memory consumption while reaching accuracy of the baseline model.
\end{itemize}

\section{RNN-Transducer}
\label{sec:rnn_transducer}

% TODO figure?
RNN-Transducer~\cite{graves2012sequence} is an architecture for automatic speech recognition (ASR).
It predicts target labels augmented with special symbol $\Blank$, which represents an alignment between a network prediction and a target sequence.
The architecture consists of three components: encoder, prediction, and joint networks.

The encoder network accepts an input speech sequence $\mathbf{x}$ and produces high-level representations $(\mathbf{h}^\mathsf{enc}_1, \cdots, \mathbf{h}^\mathsf{enc}_T)$,
where $\mathbf{h}^\mathsf{enc}_t \in \mathbb{R}^{H}$ is $H$-dimensional vector of $t$-th frame and $T$ is the number of frames.
% In this work, Conformer~\cite{gulati2020conformer} is used for the encoder network.
%
The prediction network converts previously generated output labels to high-level representation, which is used to predict the next output label.
For an output label sequence of length $u \ge 0$, the prediction network produces $\mathbf{h}^\mathsf{pre}_u \in \mathbb{R}^{H}$.
% In this work, LSTM~\cite{sepp1997long} is used for the prediction network.
%
The joint network fuses encoder representation $\mathbf{h}^\mathsf{enc}_t$ and prediction representation $\mathbf{h}^\mathsf{pre}_u$
into label logits $\mathbf{s}_{t, u} = (s_{t, u, v}: v \in \mathcal{V})$,
where $s_{t, u, v} \in \mathbb{R}$ is an unnormalized logit of a target label $v$ and
$\mathcal{V}$ is a vocabulary set augmented with $\Blank$.

By applying the softmax function to the logits, a probability distribution over $\mathcal{V}$ is induced:
\begin{equation}
\label{eq:softmax}
p_{t, u, v} = \mathsf{softmax}(\mathbf{s}_{t, u})_v
= \frac{\exp(s_{t, u, v})}{\sum_{v' \in \mathcal{V}} \exp(s_{t, u, v'})}.
\end{equation}

During decoding, RNN-Transducer maintains a state at $(t, u)$ and generates a sentence based on the state.
The model starts with initial state at $(1, 0)$, and for each state, the model draws a label $v$ with probability $p_{t, u, v}$.
If $\Blank$ is drawn, the state is updated to $(t + 1, u)$, and generation is finished when $t > T$. Otherwise, the label is appended to the output and the state is updated to $(t, u + 1)$.

\begin{figure}[t]
  \centering
  \includegraphics[width=0.9\linewidth]{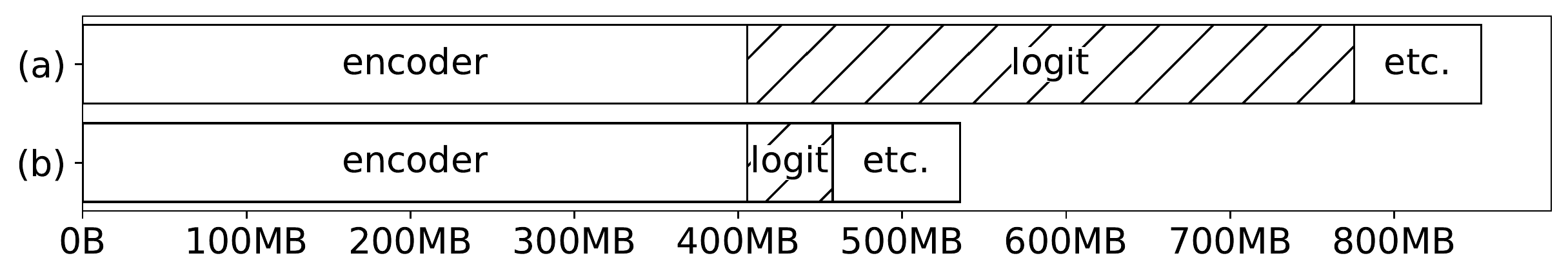}
  \vspace{-2mm}
  \caption{Memory consumption of (a) an RNN-Transducer model and (b) the same model with sampled softmax.
  % See Section~\ref{sec:experiment_librispeech} for configuration.
  % baseline vocabulary 2000 vs sampled softmax 300.
  }
  \label{fig:mem_example}
  \vspace{-6mm}
\end{figure}

During training, a target label sequence $\mathbf{y} = (y_1, \cdots, y_U)$, $y_u \in \mathcal{V} \setminus \{\Blank\}$, of length $U$ is given.
RNN-Transducer considers all possible ``alignment'', which is a state transition path compatible to $\mathbf{y}$, i.e.,\ the path generates the target $\mathbf{y}$.
Denoting $\mathcal{B}^{-1}(\mathbf{y})$ the set of alignment compatible to $\mathbf{y}$,
the probability of the target sequence is:
\begin{equation}
\label{eq:transducer_prob}
% path notation: {(t, u, y)}, y is either blank or y_u
P(\mathbf{y}|\mathbf{x}) = \sum_{\mathbf{z} \in \mathcal{B}^{-1}(\mathbf{y})} \prod_{(t, u, y) \in \mathbf{z}} p_{t, u, y}.
\end{equation}
Then the loss function is negative log-probability of the target sequence:
\begin{equation}
\label{eq:objective_transducer}
\TransducerLoss := - \log P(\mathbf{y}|\mathbf{x}).
\end{equation}
Note that $\TransducerLoss$ contains $p_{t, u, y}$ for all $t$ and $u$.
Also, due to definition of softmax in Eq.~\eqref{eq:softmax}, $p_{t, u, y}$ depends on $s_{t, u, v}$ for all $v \in \mathcal{V}$.
Thus, to compute $\TransducerLoss$ and its gradient, $O(T \cdot U \cdot |\mathcal{V}|)$ memory must be computed and stored solely for the logit,
resulting in very large memory allocation of the logit tensor.
Figure~\ref{fig:mem_example}~(a) shows estimated memory usage of an RNN-Transducer model,
where the whole encoder network and the logit tensor take nearly same memory space.
% For comparison, Connectionist Temporal Classification (CTC)~\cite{graves2006connectionist} only requires $O(T \cdot |\mathcal{V}|)$ memory for logit,
% which is negligible compared to the memory requirement of the encoder network.

\subsection{Previous works on memory consumption issue}
\label{sec:previous_works}

Huge memory consumption of training RNN-Transducer has been a serious issue,
and various methods have been proposed to overcome the issue.

One possible way is to have smaller $\mathcal{V}$ by adjusting tokenization level.
For alphabet-based languages, character-based tokenization gives smaller vocabulary than subword-level tokenization~\cite{kudo2018sentencepiece}.
% (at the cost of larger $U$)
However, \cite{he2019streaming} concludes a subword-level model outperforms a character-based model, implying large vocabulary is inevitable for better accuracy.
On the other hand, for Chinese and Japanese, character-level tokenization still gives large vocabulary, due to variety of Chinese characters.
For Mandarin, \cite{wang2021cascade} uses syllable-based tokenization to get small $\mathcal{V}$.
It requires language-dependent engineering for tokenization and an additional network for converting syllables to target characters.

It is also possible to improve network layers of RNN-Transducer for more efficient training.
\cite{wang2019exploring} and \cite{burchi2021efficient} propose encoder networks with more aggressive downsampling, reducing the number of frames ($T$).
% However, it requires careful tuning of encoder network, and may affect accuracy of the model.
\cite{saon2021advancing} proposes new architecture for a joint network using bilinear layer, and achieves better accuracy while using similar amount of memory.
% Still, training of the model remains expensive.

Improving efficiency of training has been also investigated.
\cite{li2019improving} improves memory usage of RNN-Transducer by reducing padding for the minibatch setting.
% Still, it does not change the asymptotic complexity of training.
Pre-training of RNN-Transducer networks using other objectives have been investigated, including CTC~\cite{graves2013speech,rao2017exploring}, language models~\cite{rao2017exploring}, and cross-entropy~\cite{hu2020exploring}.
% It may lead faster convergence of the model, while it is still required to train the model using RNN-Transducer objective.
In a teacher-student distillation setting, \cite{panchapagesan2021efficient} proposes to simplify distillation loss by merging non-target labels.
% while the ordinary RNN-Transducer objective remains same.

At Section~\ref{sec:sampled_softmax}, we will present a new way to reduce memory consumption by replacing $\mathcal{V}$ with much smaller sets at the training time.
The proposed method does not need any modification of tokenization, networks, or training pipeline.
Also, most of previous works above could be combined with our proposed method.

\subsection{Auxiliary CTC losses}
\label{sec:auxiliary_ctc_losses}

Auxiliary loss functions have been proposed for regularization of RNN-Transducer models~\cite{jeon2021multitask,boyer2021study,liu2020improving} and
CTC models~\cite{andros2020deja,lee2021intermediate,jumon2021relaxing}.
In this work, we consider CTC-based regularization methods for the encoder network,
as they are computationally inexpensive and will be also used for sampled softmax of Section~\ref{sec:sampled_softmax}.

Joint CTC~\cite{kim2017joint,jeon2021multitask,boyer2021study} uses encoder output $\mathbf{h}^\mathsf{enc}_t$ to compute an auxiliary CTC loss $\mathcal{L}_\mathsf{CTC} := -\log P_\mathsf{CTC}(\mathbf{y}|\mathbf{h}^\mathsf{enc}_1, \cdots, \mathbf{h}^\mathsf{enc}_T)$.

Intermediate CTC~\cite{lee2021intermediate} uses intermediate output $\mathbf{h}^\mathsf{inter}_t$ from a middle layer of an encoder network to compute an additional CTC loss $\mathcal{L}_\mathsf{InterCTC} := - \log P_\mathsf{InterCTC}(\mathbf{y}|\mathbf{h}^\mathsf{inter}_1, \cdots, \mathbf{h}^\mathsf{inter}_T)$.
% It can be jointly optimized with RNN-Transducer and joint CTC losses.
% Similar to the joint CTC loss, it takes very small cost of computation and memory.
It is similar to the auxiliary loss of~\cite{liu2020improving} which uses $\mathbf{h}^\mathsf{inter}_t$ to compute an auxiliary RNN-Transducer loss.
However, the auxiliary loss requires its own logit tensor, leading to significant memory overhead.
From preliminary experiments, we found similar improvement of two intermediate losses and decided to use intermediate CTC to save memory.

Self-conditioned CTC (SC-CTC)~\cite{jumon2021relaxing} extends intermediate CTC by conditioning the encoder using the intermediate distribution $P_\mathsf{InterCTC}$.
The distribution is combined to the intermediate output $\mathbf{h}^\mathsf{enc}_t$ then fed to the next layer of the encoder.

\section{Sampled softmax}
\label{sec:sampled_softmax}

As seen in Section~\ref{sec:rnn_transducer}, RNN-Transducer requires a significant amount of memory for training.
In Eq.~\eqref{eq:softmax}, the denominator of $p_{t, u, v}$ contains $s_{t, u, v'}$ for all $v' \in \mathcal{V}$,
which leads total memory complexity of $O(T \cdot U \cdot |\mathcal{V}|)$ for computing the objective $\TransducerLoss$ in Eq.~\eqref{eq:objective_transducer}.

To overcome the issue, we propose to apply sampled softmax~\cite{jean2015using}, originally developed for neural machine translation, to RNN-Transducer.
Sampled softmax approximates the denominator of softmax operation by choosing only a certain subset.

Let $\VocabPos = \{\Blank, y_1, \cdots, y_U\}$ a ``positive set'', which consists of labels used in alignments of Eq.~\eqref{eq:transducer_prob}.
Let $\VocabNeg \subset \mathcal{V} \setminus \VocabPos$ a ``negative set'', which is a subset sampled from some probability distributions.
The softmax in Eq.~\eqref{eq:softmax} and loss function in Eq.~\eqref{eq:objective_transducer} are modified below
so that they only depend on $\VocabSampled := \VocabPos \bigcup \VocabNeg$.

For sampled softmax, a 
probability $p^\mathsf{sampled}_{t, u, v}$ is defined for $v \in \VocabSampled$:
\begin{equation}
p^\mathsf{sampled}_{t, u, v} := \frac{\exp(s_{t, u, v})}{\sum_{v' \in \VocabSampled} \exp(s_{t, u, v'})}.
\end{equation}
Note that the denominator only considers the subset $\VocabSampled$, which makes $p^\mathsf{sampled}_{t, u, v}$ different from $p_{t, u, v}$.
Then the objective $\TransducerLoss$ in Eq.~\eqref{eq:objective_transducer} is modified by replacing $p_{t, u, y}$ with $p^\mathsf{sampled}_{t, u, y}$.

Sampled softmax approximates the original softmax by the subset $\VocabSampled$,
and the choice of $\VocabNeg$ affects the quality of approximation.
For example, if a label is rarely contained in $\VocabSampled$ during training, its probability may not be correctly learned and it may be wrongly emitted at decoding.
We sample $\VocabNeg$ independently for each minibatch;
sampling strategies and distributions are discussed at Section~\ref{sec:sampling_strategies} and Section~\ref{sec:sampling_distributions}.

With sampled softmax, it is only required to compute $s_{t, u, v'}$ for $v' \in \VocabSampled$, not the full set $\mathcal{V}$.
Thus, the memory size of a logit tensor is reduced to $O(T \cdot U \cdot |\VocabSampled|)$.
% We treat $|\VocabSampled|$ as a hyperparameter and adjust $|\mathcal{V}^\mathsf{neg}|$ depending on $|\mathcal{V}^\mathsf{pos}|$.

Figure~\ref{fig:mem_example} shows memory reduction of RNN-Transducer due to sampled softmax.
While the original model consumes nearly same memory for the encoder network and the logit tensor,
the memory of the logit tensor is reduced using sampled softmax if $|\VocabSampled|$ is sufficiently small.
Sampled softmax also allows training of RNN-Transducer with larger batch size,
which may be beneficial for certain networks like Conformer~\cite{gulati2020conformer} due to large batch dependency of Batch Normalization.

\subsection{Interaction with minibatch}
\label{sec:sampling_strategies}

\begin{figure}[t]
  \centering
  \includegraphics[width=1.0\linewidth]{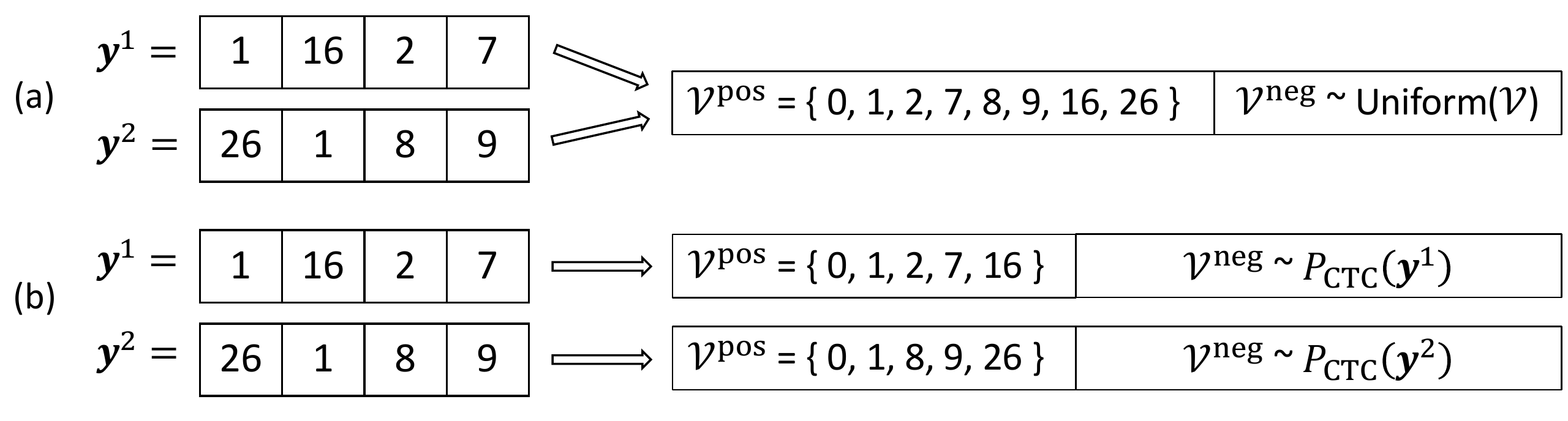}
  \vspace{-8mm}
  \caption{Comparison of (a) batched sampling and (b) example-wise sampling. Example-wise sampling has smaller $\VocabPos$ and can use different sampling distributions for $\VocabNeg$.}
  \label{fig:minibatch}
  \vspace{-6mm}
\end{figure}

For GPU-based training, multiple training examples are packed in one minibatch and computed in parallel.
This affects implementation strategy of sampled softmax.

In common implementation of sampled softmax\footnote{For example, this implementation strategy is used by TensorFlow's \texttt{sampled\_softmax\_loss}.},
$\VocabSampled$ is prepared for each minibatch,
i.e., $\VocabPos$ consists of all target labels in the minibatch and $\VocabNeg$ is shared among training examples in the minibatch,
as shown in Figure~\ref{fig:minibatch}~(a).
We call this strategy ``batched sampling''.

Batched sampling has a major drawback that $|\VocabPos|$ grows as batch size increases.
It can be significantly large (e.g., $|\VocabPos| > 500$) for large batch size, and $|\VocabSampled|$ cannot be reduced below the value, affecting memory consumption.
Also, the positive labels of other examples may not be helpful for training, leading to inefficiency of training.

Alternatively, we propose ``example-wise sampling'', which is to sample $\VocabSampled$ independently for each training example,
as shown in Figure~\ref{fig:minibatch}~(b).
This gives much smaller positive sets, and it is possible to assign a different negative set for each training example.
Especially, it is possible to use a sampling distribution conditioned on the training example.

\subsection{Sampling distributions}
\label{sec:sampling_distributions}

Sampled softmax requires a probability distribution over vocabulary for sampling $\VocabNeg$.
The simplest choice is an uniform distribution: all labels in $\mathcal{V} \setminus \VocabPos$ have equal probabilities.
However, if a sampled label is irrelevant to the training example, its probability may be already low and it may not contribute much to model accuracy.

Ideally, if a model predicts a wrong label, it should be corrected during learning.
So, it would be beneficial to sample such high-probability labels.
However, determining label probabilities would require the computation of the logit $s_{t, u, v}$ for all $v \in \mathcal{V}$,
defeating the purpose of sampled softmax.
% (we may sample some path or just select some nodes (t, u), but it would require careful and complex sampling...)

To this end, we propose to use the token posterior distribution $P_\mathsf{CTC}$ of a joint CTC loss for sampling $\VocabNeg$.
As joint CTC and RNN-Transducer share same encoder network, the joint CTC distribution would be similar to the RNN-Transducer distribution and it would be a good approximation for the RNN-Transducer model.
Similarly, we may also use the distribution $P_\mathsf{InterCTC}$ of Intermediate CTC or Self-conditioned CTC.

$P_\mathsf{CTC}$ consists of $T$ independent distributions over $\mathcal{V}$, and it needs to be transformed into a single distribution over $\mathcal{V} \setminus \VocabPos$.
We simply average the distribution over $T$ frames and assign zero probability to $\VocabPos$ to produce the desired distribution.

% For batched sampling, we also average the distribution over a minibatch to get a single distribution per minibatch. % this is removed here as it is also omitted in expr
% Similar to discussion of Section~\ref{sec:sampling_strategies}, example-wise sampling is more advantageous for model accuracy as the negative sample is not affected by other examples in minibatch.

% At Section~\ref{sec:experiments}, we show effectiveness of CTC-based sampling over uniform sampling.

\subsection{CTC-constrained decoding}
\label{sec:ctc_constrained_decoding}

When the negative set $\VocabNeg$ is drawn from joint CTC distribution as described in at Section~\ref{sec:sampling_distributions},
a label may be rarely sampled during training if its probability is very low in joint CTC\@.
If the label has a high probability from a RNN-Transducer but a low probability from a joint CTC, it may not be correctly learned and may be erroneously emitted during decoding.

To prevent such issue, we propose to constrain decoder to emit labels only in a vocabulary subset, which is determined from the distribution from the joint CTC.
% We first compute probability distribution of joint CTC branch, and determine a vocabulary subset which has high probability on the distribution.
As a simple heuristic, we average $P_\mathsf{CTC}$ over frames to get a single distribution over $\mathcal{V}$, and take top-$K$ labels based on their probabilities.
% Then, we restrict RNN-Transducer decoding to emit labels in the subset only.
This can be applied to both greedy decoding and beam-search decoding.

The proposed method can be viewed as an approximation of joint decoding of CTC and RNN-Transducer.
While the implementation of joint decoding requires complex handling of blank labels of both models~\cite{jeon2021multitask}, the proposed method can be easily applied to ordinary decoding algorithms of RNN-Transducer.

\section{Experiments}
\label{sec:experiments}

We use three corpora:
LibriSpeech~\cite{vassil2015librispeech},
AISHELL-1~\cite{bu2017aishell}, and Corpus of Spontaneous Japanese (CSJ)~\cite{kikuo2000spotaneous}.
LibriSpeech consists of read English speech; we use the full set (960h hours) for training.
AISHELL-1 consists of Mandarin speech; we use the full set (170 hours) for training.
CSJ consists of various Japanese speech including academic presentations and public speech;
we use the 271-hour subset of academic presentation speech (CSJ-APS) for training, following~\cite{yosuke2021comparative}.

For LibriSpeech, SentencePiece~\cite{kudo2018sentencepiece} tokenization is used.
SentencePiece allows user to determine vocabulary size, affecting a level of tokenization.
Small vocabulary size would be memory-efficient for RNN-Transducer, but its corresponding target sentence becomes long and may affect accuracy of the model.
To compare the effect of vocabulary size, we use various vocabulary sizes for SentencePiece from 500 to 2000.
We report word error rates (WERs) for the corpus.

For AISHELL-1 and CSJ-APS, character-level tokenization is used.
Since their transcriptions contain various Chinese and Kanji characters,
their vocabulary sizes are inevitably large: 4231 on AISHELL and 2753 on CSJ-APS\@.
We report character error rates (CERs) for the corpora.

\subsection{Results on LibriSpeech}
\label{sec:experiment_librispeech}

\begin{table}[t]
\centering
\caption{Word error rates (WERs) on LibriSpeech. No external language model is used.}
\label{table:librispeech}
\vspace{-2mm}
\scalebox{0.85}{\begin{tabular}{ccc cc}
\toprule
& & & dev & test \\
$|\mathcal{V}|$ & $|\VocabSampled|$ & AuxLoss & clean / other & clean / other \\
\midrule

\multicolumn{5}{l}{\textbf{Baseline}} \\
2000 & - & & 2.69 / 6.91  &  2.91 / 6.93   \\
2000 & - & +InterCTC & 2.63 / 6.60 & 2.84 / 6.76  \\
2000 & - & +SC-CTC & \textbf{2.51} / 6.57 &  \textbf{2.74} / \textbf{6.52} \\
600  & - &    & 2.70 / 7.16 & 2.80 / 7.17 \\
600  & - & +SC-CTC & \textbf{2.52} / 6.59 & 2.81 / 6.61 \\
500  & - &    &  2.66 / 7.12  &  2.96 / 7.25  \\
500  & - & +SC-CTC &  2.65 / 6.67  &  2.82 / 6.88  \\
\midrule
\multicolumn{5}{l}{\textbf{Batched sampling from uniform}} \\
2000 & 600 & & 2.92 / 7.23 & 3.13 / 7.30 \\
%2000 & 600 & +InterCTC &  2.76 / 6.98  & 2.93 / 7.02 \\
2000 & 600 & +SC-CTC & 2.66 / 6.80 & 2.97 / 6.88 \\
%\multicolumn{5}{l}{\textbf{%Batched sampling from SC-CTC}} \\
%2000 & 600 & +SC-CTC & 2.57 / 6.79 & 2.75 / 6.69 \\
\midrule
\multicolumn{5}{l}{\textbf{Example-wise sampling from uniform}} \\
2000 & 600 & +SC-CTC & 2.72 / 6.97 & 2.95 / 6.99 \\
500 & 250 & +SC-CTC & 2.70 / 6.81 & 3.01 / 6.93 \\
\midrule
%\multicolumn{5}{l}{\textbf{Example-wise sampling from InterCTC / SC-CTC}} \\
% todo: consider swtiching
\multicolumn{5}{l}{\textbf{Example-wise sampling from  SC-CTC}} \\
2000 & 600 & +SC-CTC & 2.58 / \textbf{6.54} & 2.82 / 6.56 \\
2000 & 300 & +SC-CTC & 2.58 / \textbf{6.53} & \textbf{2.75} / 6.63
\\
%500 & 250 & +InterCTC & 2.63 / 6.75 & 2.81 / 6.76 \\
500 & 250 & +SC-CTC   & 2.60 / 6.66 & 2.84 / 6.72 \\
500 & 160 & +SC-CTC   & 2.59 / 6.70 & 2.78 / 6.73 \\
\bottomrule
\end{tabular}
}
\vspace{-2mm} % haha
\end{table}

\begin{figure}[t]
  \centering
  \includegraphics[width=1.0\linewidth]{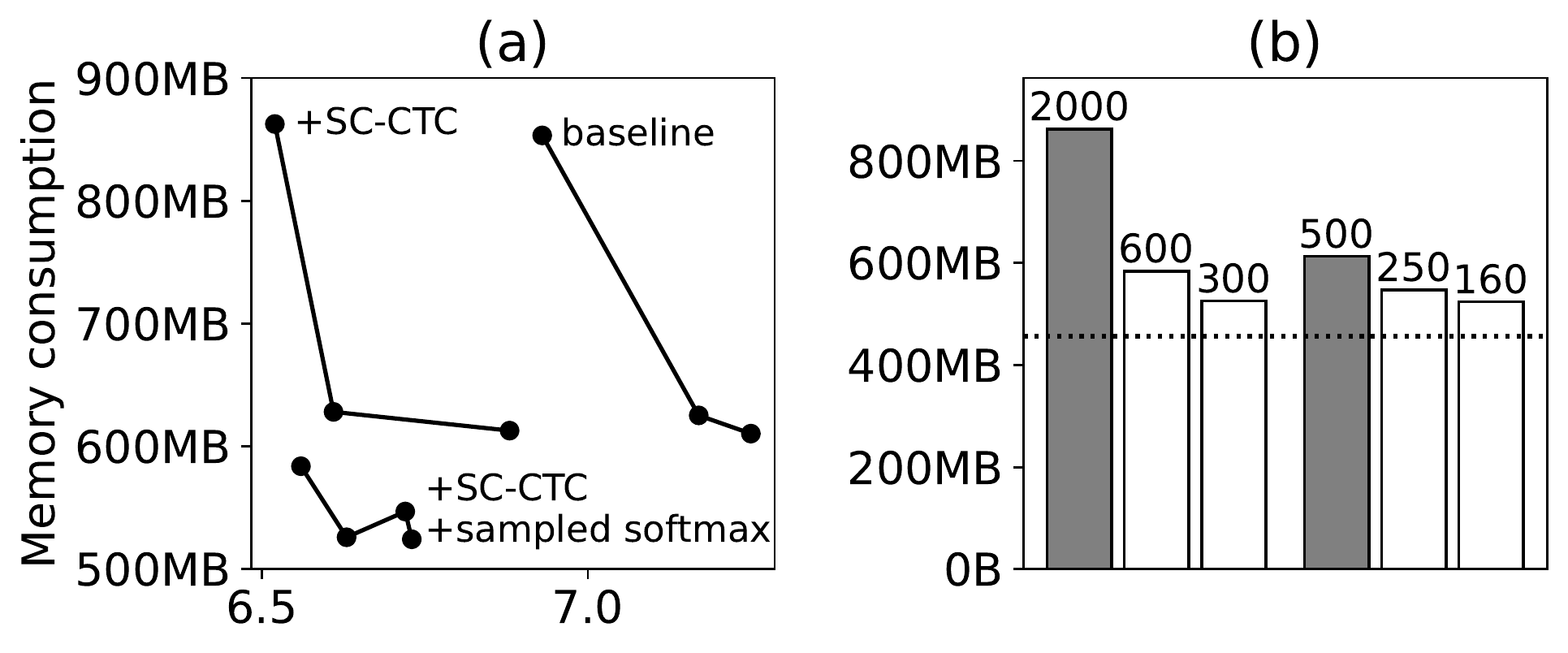}
  \vspace{-6mm}
  \caption{
  (a) Memory - test-other WER plot of LibriSpeech for various vocabulary and sample set sizes.
  (b) Memory consumption per training example for various sample set sizes.
  Gray bars are baseline models without sampled softmax.
  Dotted line is memory consumption of encoder and predictor networks.
  %Dotted line is a lower bound of sampled softmax, obtained by excluding logit tensor.
  }
  \label{fig:memplot}
  \vspace{-4mm}
\end{figure}

For LibriSpeech, we follow the architecture of Conformer-M~\cite{gulati2020conformer} for an encoder network.
SpecAugment~\cite{park2019specaugment} and speed perturbation are applied for regularization.
%and Stochastic Depth~\cite{huang2016deep}.
Beam search~\cite{graves2012sequence} is conducted for decoding and a beam size of 10 is used. No external language model is applied.
 The experimental result is reported in Table~\ref{table:librispeech}.
% the baseline accuracy is similar to values of~\cite{boyer2021study}. (which uses layer=12 and ff=2048, while we use layer=16 and ff=1024...)

For the choice of a sampling distribution of sampled softmax, we found using the distribution of Self-conditioned CTC yields better accuracy than using the uniform distribution does.
Also, we found example-wise sampling gives better accuracy than batched sampling does, and allows smaller $\VocabSampled$ for training.
The models trained with example-wise sampling from Self-conditioned CTC reach or even outperform the baseline models with Self-conditioned CTC.

Figure~\ref{fig:memplot}~(a) shows memory consumption and test-other WER for various configurations of $|\mathcal{V}|$ and $|\VocabSampled|$.
Self-conditioned CTC improves baseline models with marginal memory consumption,
and sampled softmax greatly reduces memory consumption with little or no degradation of accuracy.

We found using large vocabulary generally leads to better accuracy, as shown in Figure~\ref{fig:memplot}~(a).
This shows the advantage of large vocabulary with sampled softmax over small vocabulary.
Interestingly, the model with $|\mathcal{V}|=2000$ and $|\VocabSampled|=600$ consumes less memory than the baseline model of $|\mathcal{V}|=500$, because larger vocabulary leads to shorter target sequences.

Figure~\ref{fig:memplot}~(b) shows memory consumption of various configurations of $\mathcal{V}$ and $\VocabSampled$.
The baseline model of $|\mathcal{V}| = 2000$ takes nearly 850MB memory per training example,
and the sampled softmax with $|\VocabSampled|=300$ only takes 530MB memory, and the amount of a logit tensor is now much smaller than the amount of encoder and predictor networks, displayed as a dotted line in Figure~\ref{fig:memplot}~(b).

\subsection{Results on AISHELL-1 and CSJ-APS}

\begin{table}[t]
\centering
\caption{Character error rates (CERs) on AISHELL-1 and CSJ-APS\@. No external language model is used.}
\label{table:aishell_csj}
% set \scalebox{0.9}{ ... } if table is too big
\vspace{-2mm}
\scalebox{0.85}{
\begin{tabular}{cc cc c}
\toprule

&& AISHELL-1  & CSJ-APS  \\
&& {($|\mathcal{V}|=4231$)} & {($|\mathcal{V}|=2753$)} \\

\cmidrule(l{0.3em}r{0.3em}){3-3} \cmidrule(l{0.3em}r{0.3em}){4-4}
$|\VocabSampled|$&AuxLoss& dev / test & eval1 / eval2 / eval3 \\

\midrule
\multicolumn{4}{l}{\textbf{Baseline}} \\
- &           & 4.4 / 4.8 & 5.3 / {3.9} / 9.5 \\
- & +InterCTC & \textbf{4.2} / 4.7 & 5.3 / {3.9} / \textbf{9.0} \\
\midrule
\multicolumn{4}{l}{\textbf{Batched sampling from uniform}} \\
500  & & 4.4 / 4.9 & 5.5 / 4.0 / 9.4 \\
1000 & & 4.4 / 4.9 & 5.4 / {3.9} / 9.5 \\
\midrule
\multicolumn{4}{l}{\textbf{Example-wise sampling from uniform}} \\
500  & & 4.5 / 5.0 & {5.2} / 4.0 / 9.5 \\
1000 & & 4.4 / 4.9 & 5.4 / {3.9} / 9.4 \\
\midrule
\multicolumn{4}{l}{\textbf{Example-wise sampling from joint-CTC}} \\
500  &              & 4.4 / 4.9 & 5.3 / 4.0 / 9.5 \\
500  & +InterCTC    & \textbf{4.2} / \textbf{4.6} & {5.2} / {3.9} / 9.3 \\
500  & +large batch & - & \textbf{5.1} / \textbf{3.7} / 9.1 \\
1000 &              & 4.4 / 4.8 & 5.5 / {3.9} / 9.4 \\

\bottomrule
\end{tabular}
}
\vspace{-6mm} % haha
\end{table}

For AISHELL-1 and CSJ-APS, we follow the standard recipe~\cite{boyer2021study} of ESPnet~\cite{watanabe2018espnet}, which uses 12-layer Conformer~\cite{gulati2020conformer} for the encoder network and 1-layer LSTM~\cite{sepp1997long} for the prediction network.
% ff dim 2048, total 83M parameters
SpecAugment~\cite{park2019specaugment}, speed perturbation, and joint CTC loss are applied for regularization.
% For sampled softmax, we use 500 and 1000 samples for $\VocabSampled$.
% All experiment used same batch size for comparison unless noted.
For decoding, we use CTC-constrained greedy decoding with top-100 labels as described in Section~\ref{sec:ctc_constrained_decoding}. % as beam search gives only marginal improvements.
Table~\ref{table:aishell_csj} shows character error rates (CERs) on AISHELL-1 and CSJ-APS\@.

Similar to Section~\ref{sec:experiment_librispeech}, sampling from the joint CTC distribution yields better accuracy than sampling from the uniform distribution, although the difference is small here.
We found sampling from joint-CTC with $|\VocabSampled| = 500$ is sufficient to match the baseline.
The size of sample set is only 12\% and 18\% of vocabulary sets on AISHELL-1 and CSJ-APS, respectively.
% This gives significant memory reduction as shown in Figure~\ref{fig:memplot}~(c).

For AISHELL-1, the model trained with sampled softmax and Intermediate CTC achieves the lowest CER in Table~\ref{table:aishell_csj}, and even outperforms the previous result in~\cite{boyer2021study}.
Also, for CSJ-APS, it is possible to increase batch size with sampled softmax due to memory reduction.
With large batch size, accuracy of the model is improved and the CERs are comparable to the state-of-the-art result in~\cite{yosuke2021comparative}.
This illustrates the importance of memory reduction for training RNN-Transducer models.

\section{Conclusions}

We propose to apply sampled softmax to RNN-Transducer for memory-efficient training.
During training, a subset of vocabulary is sampled for each iteration, which leads to great memory reduction.
We propose two extensions of sampled softmax:
the example-wise sampling strategy for efficient implementation for minibatch setting,
and the employment of joint CTC and Self-conditioned CTC for a sampling distribution of the subset.
We experimentally show that sampled softmax gives huge memory reduction while achieving the accuracy of the baseline model.
%We also show Intermediate CTC and Self-conditioned CTC further reduces error rates of RNN-Transducer model.

\bibliographystyle{IEEEtran}
\bibliography{bib}

\end{document}